\documentclass[10pt,twocolumn,journal]{IEEEtran}
\usepackage{times}
\usepackage[final]{graphicx}
\usepackage[reqno]{amsmath}
\usepackage{amsfonts}
\usepackage{times,amsmath,epsfig}
\usepackage{latexsym,amssymb}
\usepackage{cite}

\def\myQED{\mbox{\rule[0pt]{1.5ex}{1.5ex}}}

\newcommand{\thetav}{\hbox{\boldmath$\theta$}}

\newtheorem{rl}{Rule}

\newtheorem{lem}{Lemma}

\newtheorem{exmpl}{Example}
\newcommand{\qed}{\hfill $\Box$}
\newcommand{\no}{\nonumber}

\begin{document}

\title{Cognitive Medium Access: Exploration, Exploitation and Competition}
\author{
Lifeng Lai, Hesham El Gamal, Hai Jiang and H. Vincent Poor
\thanks{L. Lai and H. V. Poor (\{llai,poor\}@princeton.edu) are
with the Department of Electrical Engineering at Princeton
University. H. El Gamal (helgamal@ece.osu.edu) is with the
Department of Electrical and Computer Engineering at the Ohio
State University and is currently visiting Nile University, Cairo,
Egypt. H. Jiang (hai.jiang@ece.ualberta.ca) is with the Department
of Electrical and Computer Engineering at the University of
Alberta. This research was supported by the National Science
Foundation under Grants ANI-03-38807 and CNS-06-25637.}}
\maketitle 



\begin{abstract}
This paper establishes the equivalence between cognitive medium
access and the competitive multi-armed bandit problem. First, the
scenario in which a single cognitive user wishes to
opportunistically exploit the availability of empty frequency
bands in the spectrum with multiple bands is considered. In this
scenario, the availability probability of each channel is unknown
to the cognitive user a priori. Hence efficient medium access
strategies must strike a balance between exploring the
availability of other free channels and exploiting the
opportunities identified thus far. By adopting a Bayesian approach
for this classical bandit problem, the optimal medium access
strategy is derived and its underlying recursive structure is
illustrated via examples. To avoid the prohibitive computational
complexity of the optimal strategy, a low complexity
asymptotically optimal strategy is developed. The proposed
strategy does not require any prior statistical knowledge about
the traffic pattern on the different channels. Next, the
multi-cognitive user scenario is considered and low complexity
medium access protocols, which strike the optimal balance between
exploration and exploitation in such competitive environments, are
developed. Finally, this formalism is extended to the case in
which each cognitive user is capable of sensing and using multiple
channels simultaneously.
\end{abstract}

\section{Introduction} \label{sec:intro}
Recently, the opportunistic spectrum access problem has been the
focus of significant research
activity~\cite{Mitola:IPC:99,Haykin:JSAC:05,Devroye:TIT:06}. The
underlying idea is to allow unlicensed users (i.e., cognitive
users) to access the available spectrum when the licensed users
(i.e., primary users) are not active. The presence of high
priority primary users and the requirement that the cognitive
users should not interfere with them define a new medium access
paradigm which we refer to as \emph{cognitive medium access}. The
overarching goal of our work is to develop a unified framework for
the design of efficient, and low complexity, cognitive medium
access protocols.

The spectral opportunities available to the cognitive users are
expected to be time-varying on different time-scales. For example,
on a small scale, multimedia data traffic of the primary users
will tend to be bursty~\cite{Sahinoglu:ICM:99}. On a large scale,
one would expect the activities of each user to vary throughout
the day. Therefore, to avoid interfering with the primary network,
the cognitive users must first probe to determine whether there
are primary activities in each channel before transmission. Under
the assumption that each cognitive user cannot access all of the
available channels simultaneously, the main task of the medium
access protocol is to distributively choose which channels each
cognitive user should attempt to use in different time slots, in
order to fully (or maximally) utilize the spectral opportunities.
This decision process can be enhanced by taking into account any
available statistical information about the primary traffic. For
example, with a single cognitive user capable of accessing
(sensing) only one channel at a time, the problem becomes trivial
if the probability that each channel is free is known {\em a
priori}. In this case, the optimal rule is for the cognitive user
to access the channel with the highest probability of being free
in all time slots. However, such time-varying traffic information
is typically not available to the cognitive users {\em a priori}.
The need to learn this information on-line creates a fundamental
tradeoff between exploitation and exploration. Exploitation refers
to the short-term gain resulting from accessing the channel with
the estimated highest probability of being free (based on the
results of previous sensing decisions) whereas exploration is the
process by which the cognitive user learns the statistical
behavior of the primary traffic (by choosing possibly different
channels to probe across time slots). In the presence of multiple
cognitive users, the medium access algorithm must also account for
the competition between different users over the same channel.

In this paper, we develop a unified framework for the design and
analysis of cognitive medium access protocols. As argued in the
sequel, this framework allows for the construction of strategies
that strike an optimal balance among exploration, exploitation and
competition.  The key observation motivating our approach is the
equivalence between our problem and the classical multi-armed
bandit problem (see~\cite{Berry:Book:85} and references therein).
This equivalence allows for building a solid foundation for
cognitive medium access using tools from reinforcement machine
learning~\cite{Sutton:Book:98}. The connection between cognitive
medium access and the multi-armed bandit problem has been
independently and concurrently observed
in~\cite{Motamedi:ETTRT:07}. That work, however, is limited to
special cases of the general approach presented here. In
particular, in~\cite{Motamedi:ETTRT:07}, the channels are assumed
to be independent and the goal is to maximize the discounted sum
of throughput, which is the problem addressed in
Example~\ref{exmpl:ind} in Section~\ref{sec:full} below. A related
work also appears in~\cite{Zhao:JSAC:07}, in which the
availability of each channel is assumed to follow a Markov chain,
whose transition matrix is known to the cognitive user. The only
uncertainty faced by the cognitive user in that work is the
particular realization of the channel, while in our work the
cognitive users also need to learn the statistics of the channel
in real time.

We consider three scenarios in this paper. In the first scenario,
we assume the existence of a single cognitive user capable of
accessing only a single channel at any given time. In this
setting, we derive an optimal sensing rule that maximizes the
expected throughput obtained by the cognitive user. Compared with
a genie-aided scheme, in which the cognitive user knows {\em a
priori} the primary network traffic information, there is a
throughput loss suffered by any medium access strategy. We obtain
a lower bound on this loss and further construct a linear
complexity single index protocol that achieves this lower bound
asymptotically (when the primary traffic behavior changes very
slowly). In the second scenario, we design distributed sensing
rules that account for the competitive dimension of the problem in
which the cognitive users must also take the competition from
other cognitive users into consideration when making sensing
decisions. We first characterize the optimal distributed sensing
rule for the case in which the traffic information of the primary
network is available to the cognitive users. Under this idealistic
assumption, we show that the throughput loss of the proposed
distributed sensing rule, compared with a throughput optimal
centralized scheme, goes to zero exponentially as the number of
cognitive users increases. To prevent any possible misbehavior by
the cognitive users, we further design a game theoretically fair
sensing rule, whose loss compared with the throughput optimal
centralized rule also goes to zero exponentially. Building on
these results, we then devise distributed sensing rules that do
not require prior knowledge about the traffic and converge to the
optimal distributed rule and game theoretically fair rule,
respectively. In the third scenario, we extend our work to the
case in which the cognitive user is capable of accessing more than
one channel simultaneously.

The rest of the paper is organized as follows. Our network model
is detailed in Section~\ref{sec:model}. Section~\ref{sec:full}
analyzes the scenario in which a single cognitive user capable of
sensing one channel at a time is present. The extension to the multi-user case is reported in Section~\ref{sec:mulp} whereas the
multi-channel extension is studied in
Section~\ref{sec:multisensing}. Finally, Section~\ref{sec:con}
summarizes our conclusions.

\section{Network Model}\label{sec:model}

Throughout this paper, upper-case letters (e.g.,  $X$) denote
random variables, lower-case letters (e.g., $x$)  denote
realizations of the corresponding random variables, and
calligraphic letters (e.g,  $\mathcal{X}$) denote  finite alphabet
sets over which corresponding variables range. Also, upper-case
boldface letters (e.g.,  $\mathbf{X}$) denote random vectors and
lower-case boldface letters (e.g., $\mathbf{x}$) denote
realizations of the corresponding random vectors.

\begin{figure}[thb]
\centering
\includegraphics[width=0.45 \textwidth]{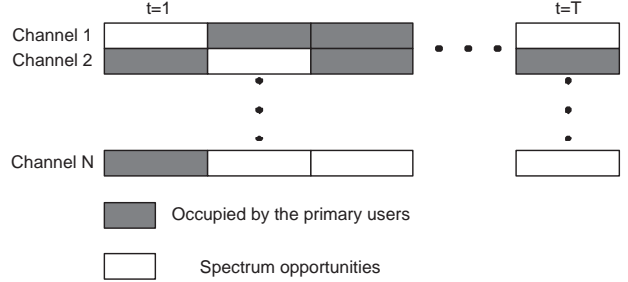}
\caption{Channel model.} \label{fig:model}
\end{figure}

Figure~\ref{fig:model} shows the channel model of interest. We
consider a primary network consisting of $N$ channels,
$\mathcal{N}=\{1,\cdots,N\}$, each with bandwidth $B$. The users
in the primary network are operated in a synchronous time-slotted
fashion. We use $i$ to refer to the channel index, $j$ to refer to
the time-slot index and $k$ referring to the index of the
cognitive users. We assume that at each time slot, channel $i$ is
free with probability $\theta_i$. Let $Z_{i}(j)$ be a random
variable that equals $1$ if channel $i$ is free at time slot $j$
and equals $0$ otherwise. Hence, given $\theta_i$, $Z_{i}(j)$ is a
Bernoulli random variable with probability density function (pdf)
$$h_{\theta_i}(z_{i}(j))=\theta_i\delta(1)+(1-\theta_i)\delta(0),$$
where $\delta(\cdot)$ is the delta function. Furthermore, for a
given $\thetav=(\theta_1,\cdots,\theta_N)$, $Z_{i}(j)$ are
independent for each $i$ and $j$. We consider a block varying
model in which the value of $\thetav$ is fixed for a block of $T$
time slots and randomly changes at the beginning of the next block
according to some joint pdf $f(\thetav)$. Our results can also be
extended to the scenarios in which $Z_i(j)$s follow a Markov chain
model.

In our model, the cognitive users attempt to exploit the
availability of free channels in the primary network by sensing
the activity at the beginning of each time slot. Our work seeks to
characterize efficient strategies for choosing which channels to
sense (access). The challenge here stems from the fact that the
cognitive users are assumed to be unaware of $\thetav$ {\em a
priori}.  We consider two cases in which the cognitive user either
has or does not have prior information about the pdf of $\thetav$,
i.e., $f(\thetav)$. To further illustrate the point, let us
consider our first scenario in which a single cognitive user
capable of sensing only one channel is present. At time slot $j$,
the cognitive user selects one channel $S(j)\in\mathcal{N}$ to
access. If the sensing result shows that channel $S(j)$ is free,
i.e., $Z_{S(j)}(j)=1$, the cognitive user can send $B$ bits over
this channel; otherwise, the cognitive user will wait until the
next time slot and pick a possibly different channel to access
(throughout the paper, it is assumed that the outcome of the
sensing algorithm is error free). Therefore, the total number of
bits that the cognitive user is able to send over one block (of
$T$ time slots) is
\begin{eqnarray}
W=\sum\limits_{j=1}^{T}BZ_{S(j)}(j).\no
\end{eqnarray}

It is now clear that $W$ is a random variable that depends on the
traffic in the primary network and, more importantly for us, on
the medium access protocols employed by the cognitive user.
Therefore, the overarching goal of Section~\ref{sec:full} is to
construct low complexity medium access protocols that maximize
\begin{eqnarray}
\mathbb{E}\{W\}=\mathbb{E}\left\{\sum
\limits_{j=1}^{T}BZ_{S(j)}(j)\right\}.
\end{eqnarray}

Intuitively, the cognitive user would like to select that channel
with the highest probability of being free in order to obtain more
transmission opportunities. If $\thetav$ is known then this
problem is trivial: the cognitive user should choose the channel
$i^*=\arg\max\limits_{i\in\mathcal{N}}\theta_i$ to sense. The
uncertainty in $\thetav$ imposes a fundamental tradeoff between
exploration, in order to learn $\thetav$, and exploitation, by
accessing the channel with the highest estimated free probability
based on current available information, as detailed in the
following sections.

\section{Single User--Single Channel}\label{sec:full}
We start by developing the optimal solution to the single
user--single channel scenario under the idealized assumption that
$f(\thetav)$ is known {\em a priori} by the cognitive user. As
argued next, the optimal medium access algorithm suffers from a
prohibitive computational complexity that grows exponentially with
the block length $T$. This motivates the design of low complexity
asymptotically optimal approaches that are considered next.
Interestingly, the proposed low complexity technique does not
require prior knowledge about $f(\thetav)$.
\subsection{Bayesian Approach}\label{sec:opgen}

Our single user--single channel cognitive medium access problem
belongs to the class of bandit problems. In this setting, the
decision maker must sequentially choose one process to observe
from $N\geq 2$ stochastic processes. These processes usually have
parameters that are unknown to the decision maker and, associated
with each observation is a utility function. The objective of the
decision maker is to maximize the sum or discounted sum of the
utilities via a strategy that specifies which process to observe
for every possible history of selections and observations. The
following classical example illustrates the challenge facing our
decision maker: A gambler enters a casino having $N$ slot
machines, the $i^{th}$ of which has winning probability
$\theta_i,i\in\mathcal{N}$. The gambler does not know the values
of the $\theta_i$s and must sequentially chooses machines to play.
The goal is to maximize the overall gain for a total of $T$ plays.
In this example, the stochastic processes are the outcomes of the
slot machines, the utility function is the reward that the gambler
gains each time and the gambling strategy specifies which machine
to play based on each possible past information pattern. A
comprehensive treatment covering different variants of bandit
problems can be found in~\cite{Berry:Book:85}.


We are now ready to rigorously formulate our problem. The
cognitive user employs a medium access strategy $\Gamma$, which
will select channel $S(j)\in\mathcal{N}$ to sense at time slot $j$
for any possible causal information pattern obtained through the
previous $j-1$ observations:
$$\Psi(j)=\{s(1),z_{s(1)}(1),\cdots,s(j-1),z_{s(j-1)}(j-1)\}, j\geq 2,$$
i.e. $s(j)=\Gamma(f,\Psi(j))$. Notice that $z_{s(j)}(j)$ is the
sensing outcome of the $j$th time slot, in which $s(j)$ is the
channel being accessed. If $j=1$, there is no accumulated
information, thus $\Psi(1)=\phi$ and $s(1)=\Gamma(f)$. $\Gamma$
could be stochastic, i.e., for certain $\Psi(j)$, the cognitive
user may randomly pick channel $i$ from a set
$\mathcal{A}\subseteq\mathcal{N}$ with probability $p_i$, such
that $\sum\limits_{i\in \mathcal{A}}p_i=1$. The utility that the
cognitive user obtains by making decision $S(j)$ at time slot $j$
is the number of bits it can transmit at time slot $j$, which is
$BZ_{S(j)}(j)$. We denote the expected value of the payoff
obtained by a cognitive user who uses strategy $\Gamma$ as
\begin{eqnarray}\label{eq:prob}
W_{\Gamma}=\mathbb{E}_{f}\left\{\sum
\limits_{j=1}^{T}BZ_{S(j)}(j)\right\}.
\end{eqnarray}

We denote $V^*(f,T)=\sup\limits_{\Gamma}W_{\Gamma}$, which is the
largest throughput that the cognitive user could obtain when the
spectral opportunities are governed by $f(\thetav)$ and the exact
value of each realization of $\thetav$ is not known by the user.

Each medium access decision made by the cognitive user has two
effects. The first one is the short term gain, i.e., an immediate
transmission opportunity if the chosen channel is found free. The
second one is the long term gain, i.e., the updated statistical
information about $f(\thetav)$. This information will help the
cognitive user in making better decisions in the future stages.
There is an interesting tradeoff between the short and long term
gains. If we only want to maximize the short term gain, we can
pick the one with the highest free probability to sense, based on
the current information. This myopic strategy maximally exploits
the existing information. On the other hand, by picking other
channels to sense, we gain valuable statistical information about
$f(\thetav)$ which can effectively guide future decisions. This
process is typically referred to as exploration.

More specifically, let $f^{j}(\thetav)$ be the updated pdf after
making $j-1$ observations. We begin with
$f^{1}(\thetav)=f(\thetav)$. After observing $z_{s(j)}(j)$, we
update the pdf using the following Bayesian formula.
\begin{enumerate}

\item If $z_{s(j)}(j)=1$
\begin{eqnarray}\label{eq:up1}
f^{j+1}(\thetav)=\frac{\theta_{s(j)}f^{j}(\thetav)}{\int\theta_{s(j)}f^j(\thetav)d\thetav},
\end{eqnarray}
\item If $z_{s(j)}(j)=0$
\begin{eqnarray}\label{eq:up2}
f^{j+1}(\thetav)=\frac{\left(1-\theta_{s(j)}\right)f^{j}(\thetav)}{\int\left(1-\theta_{s(j)}\right)f^j(\thetav)d\thetav}.
\end{eqnarray}
\end{enumerate}
Now, lemma 2.3.1 of~\cite{Berry:Book:85} proves that every bandit
problem with finite horizon has an optimal solution. Applying this
result to our set-up, we obtain the following.

\begin{lem}\label{lem:optimal}
For any prior pdf $f$, there exists an optimal strategy $\Gamma^*$
to the channel selection problem~\eqref{eq:prob}, and $V^*(f,T)$
is achievable. Moreover, $V^*$ satisfies the following condition:
\begin{eqnarray}\label{eq:solution}
V^*(f,T)=\max\limits_{s(1)\in\mathcal{N}}\mathbb{E}_{f}\left\{BZ_{s(1)}+V^*\left(f_{Z_{s(1)}},T-1\right)\right\},
\end{eqnarray}
where $f_{Z_{s(1)}}$ is the conditional pdf updated
using~\eqref{eq:up1} and~\eqref{eq:up2} as if the cognitive user
chooses $s(1)$ and observes $Z_{s(1)}$. Also,
$V^*\left(f_{Z_{s(1)}},T-1\right)$ is the value of a bandit
problem with prior information $f_{Z_{s(1)}}$ and $T-1$ sequential
observations.\qed
\end{lem}

In principle, Lemma~\ref{lem:optimal} provides the solution to
problem~\eqref{eq:prob}. Effectively, it decouples the calculation
at each stage, and hence, allows the use of dynamic programming to
solve the problem. The idea is to solve the channel selection
problem with a smaller dimension first and then use backward
deduction to obtain the optimal solution for a problem with a
larger dimension. Starting with $T=1$, the second term inside the
expectation in~\eqref{eq:solution} is 0, since $T-1=0$. Hence, the
optimal solution is to choose channel $i$ with the largest
$\mathbb{E}_f\{BZ_{i}\}$, which can be calculated as
\begin{eqnarray}
\mathbb{E}_f\{BZ_{i}\}=B\int \theta_i f(\thetav)d\thetav.\no
\end{eqnarray}
And $V^*(f,1)=\max\limits_{i\in\mathcal{N}}\mathbb{E}_f\{BZ_i\}$.

With the solution for $T=1$ at hand, we can now solve the $T=2$
case using~\eqref{eq:solution}. At first, for every possible
choice of $s(1)$ and possible observation $z_{s(1)}$, we calculate
the updated pdf $f_{z_{s(1)}}$ using~\eqref{eq:up1} and
\eqref{eq:up2}. Next, we calculate $V^*(f_{z_{s(1)}},1)$ (which is
equivalent to the $T=1$ problem described above). Finally,
applying~\eqref{eq:solution}, we have the following equation for
the channel selection problem with $T=2$
\begin{eqnarray}
V^*(f,2)&=&\max\limits_{i\in \mathcal{N}}\int
\left[B\theta_i+\theta_i
V^{*}(f_{z_i=1},1)\right.\no\\&&\hspace{12mm}\left.+(1-\theta_i)V^*(f_{z_i=0},1)\right]f(\thetav)d\thetav.\no
\end{eqnarray}
Correspondingly, the optimal solution is
$\Gamma^*(f)=\arg\max\limits_{i\in\mathcal{N}}V^*(f,2)$, i.e., in
the first step, the cognitive user should choose
$i^*(1)=\arg\max\limits_{i\in\mathcal{N}}V^*(f,2)$ to sense. After
observing $z_{i^*(1)}$, the cognitive user has
$\Psi(1)=\{z_{i^*(1)}\}$, and it should choose
$i^*(2)=\arg\max\limits_{i\in\mathcal{N}}V^*(f_{z_{i^*(1)}},1)$
implying that
$\Gamma^*(f,\Psi(1))=\arg\max\limits_{i\in\mathcal{N}}V^*(f_{z_{i^*(1)}},1)$.

Similarly, after solving the $T=2$ problem, one can proceed to
solve the $T=3$ case. Using this procedure recursively, we can
solve the problem with $T-1$ observations. Finally, our original
problem with $T$ observations is solved as follows.
\begin{eqnarray}
V^*(f,T)&=&\max\limits_{i\in \mathcal{N}}\int
\left[B\theta_i+\theta_i V^{*}(f_{z_i=1},T-1)\right.\no\\
&&\hspace{8mm}\left.+(1-\theta_i)V^*(f_{z_i=0},T-1)\right]f(\thetav)d\thetav.\no
\end{eqnarray}

\begin{exmpl}\label{exmpl:1}
Suppose we have two channels and two observations per block, i.e.,
$\mathcal{N}=\{1,2\}$ and $T=2$. The channels are known to be
either both very busy or both relatively idle which is reflected
in the following joint pdf
\begin{eqnarray}
f(\theta_1,\theta_2)=\frac{4}{5}\delta(0.1,0)+\frac{1}{5}\delta(0.8,1),\no
\end{eqnarray}
where $\delta(x,y)$ is the delta function at point $(x,y)$. For
simplicity of presentation, we assume that $B=100$.

In this example, on the average, channel $1$ is available with
probability $4/5\times0.1+1/5\times0.8=0.24$, whereas channel $2$
is available with probability $4/5\times0+1/5\times1=0.2$. Hence,
if the cognitive user ignores the information gained from sensing,
it should always choose channel $1$ to sense, resulting in an
average throughput of $2\times0.24\times100=48$ bits per block.
Now, we use the procedure described above to derive the optimal
rule and corresponding throughput.

1) First calculate all possible updated pdf after one
step.

If $s(1)=1,z_{s(1)}=1$, we have
\begin{eqnarray}
&&\hspace{-11mm}P(\theta_1=0.1,\theta_2=0|z_{s(1)}=1)\no\\
&&=\frac{P(z_1=1|\theta_1=0.1,\theta_2=0)P(\theta_1=0.1,\theta_2=0)}{P(z_1=1)}\no\\
&&=\frac{0.1\times 0.8}{0.8\times 0.1+0.2\times
0.8}=\frac{1}{3}.\no
\end{eqnarray}
Hence, for this case, we have
\begin{eqnarray}
f_{\{s(1)=1,z_{s(1)}=1\}}=\frac{1}{3}\delta(0.1,0)+\frac{2}{3}\delta(0.8,1).\no
\end{eqnarray}
Similarly, we obtain the following updated pdf
\begin{eqnarray}
f_{\{s(1)=1,z_{s(1)}=0\}}&=&\frac{18}{19}\delta(0.1,0)+\frac{1}{19}\delta(0.8,1),\no\\
f_{\{s(1)=2,z_{s(1)}=1\}}&=&\delta(0.8,1),\no\\
f_{\{s(1)=2,z_{s(1)}=0\}}&=&\delta(0.1,0).\no
\end{eqnarray}

2) With the updated distribution information, we solve four
channel-selection problems with $T=1$. For example, with
$f_{\{s(1)=1,z_{s(1)}=1\}}=\frac{1}{3}\delta(0.1,0)+\frac{2}{3}\delta(0.8,1)$,
if the cognitive user choose channel 1, the expected payoff would
be
$$100\times\left(\frac{1}{3}\times0.1+\frac{2}{3}\times
0.8\right)=\frac{170}{3}.$$ If the cognitive user choose channel
$2$, the expected payoff would be
$$100\times\left(\frac{1}{3}\times0+\frac{2}{3}\times
1\right)=\frac{200}{3}.$$ Thus
$$V^*(f_{\{s(1)=1,z_{s(1)}=1\}},1)=\max\{170/3,200/3\}=200/3,$$ and
the user should choose channel $2$.

Similarly, we have
$$V^*(f_{\{s(1)=1,z_{s(1)}=0\}},1)=100\times\max\{26/190,1/19\}=260/19,$$ and
the user should choose channel $1$.
$$V^*(f_{\{s(1)=2,z_{s(1)}=1\}},1)=\max\{80,100\}=100,$$ and
the user should choose channel $2$.
$$V^*(f_{\{s(1)=2,z_{s(1)}=0\}},1)=\max\{10,0\}=10,$$ and
the user should choose channel $1$.

3) Finally, we solve the problem with pdf $f$ and $T=2$.
If the cognitive user chooses channel $1$ in the first step, we
calculate
\begin{eqnarray}
&&\hspace{-6mm}\mathbb{E}_{f}\{BZ_{1}+V^*(f_{Z_1},1)\}\no\\&&\hspace{-3mm}=P(\theta_1=0.1)\Big[100\times0.1+0.1\times
V^*(f_{\{s(1)=1,z_{s(1)}=1\}},1)\no\\ &&\hspace{10mm}+(1-0.1)\times V^*(f_{\{s(1)=1,z_{s(1)}=0\}},1)\Big]\no\\
&&+P(\theta_1=0.8)\Big[100\times0.8+0.8\times
V^*(f_{\{s(1)=1,z_{s(1)}=1\}},1)\no\\&&\hspace{10mm}+(1-0.8)\times V^*(f_{\{s(1)=1,z_{s(1)}=0\}},1)\Big]\no\\
&&=252/5.\no
\end{eqnarray}
Similarly, if the cognitive user chooses channel $2$ in the first
step, we calculate
\begin{eqnarray}
&&\mathbb{E}_{f}\{BZ_{2}+V^*(f_{Z_2},1)\}\no\\&&=P(\theta_2=0)\Big[100\times0+0\times
V^*(f_{\{s(1)=2,z_{s(1)}=1\}},1)\no \\ &&+V^*(f_{\{s(1)=2,z_{s(1)}=0\}},1)\Big]\no\\
&&+P(\theta_2=1)\Big[100+V^*(f_{\{s(1)=2,z_{s(1)}=1\}},1)\no \\ &&+(1-1)V^*(f_{\{s(1)=0,z_{s(1)}=0\}},1)\Big]\no\\
&&=P(\theta_2=0)V^*(f_{\{s(1)=2,z_{s(1)}=0\}},1)\no \\
&&+P(\theta_2=1)[100+V^*(f_{\{s(1)=2,z_{s(1)}=1\}},1)]=\frac{240}{5}.\no
\end{eqnarray}

Thus
\begin{eqnarray}
V^*(f,2)&=&\max\limits_{s(1)\in\mathcal{N}}\mathbb{E}_{f}\left\{BZ_{s(1)}+V^*\left(f_{Z_{s(1)}},1\right)\right\}\no
\\ &&=\max\{252/5,240/5\}=252/5.\no
\end{eqnarray}
Hence, the optimal strategy is $\Gamma^*(f)=1$,
$\Gamma^*(f,z_1=1)=2$, $\Gamma^*(f,z_1=0)=1$. In other words, the
cognitive user should sense channel $1$ in the first time slot.
Interestingly, if channel $1$ is found free, the user should
switch to channel $2$ in the second time slot. On the other hand,
if channel $1$ is found busy, the cognitive user should keep
sensing channel $1$ at the second time slot. Finally, we observe
that the optimal strategy offers a gain of $12/5$ bits, on
average, as compared with the myopic strategy.\qed
\end{exmpl}

The optimal solution presented above can be simplified when
$f(\thetav)$ has a certain structure, as illustrated by the
following examples.

\begin{exmpl} (Symmetric Channels)
We have $N=2$ channels. Without loss of generality, let $0\leq
\theta_b<\theta_a\leq 1$. At any block, either 1) channel $1$ has
probability $\theta_a$ of being free and channel $2$ has
probability $\theta_b$ of being free or 2) channel $1$ has
probability $\theta_b$ of being free and channel $2$ has
probability $\theta_a$ of being free. The cognitive user does not
know exactly which case happens. The prior pdf information is thus
given by
\begin{eqnarray}
f(\theta_1,\theta_2)=\xi\delta(\theta_a,\theta_b)+(1-\xi)\delta(\theta_b,\theta_a),\no
\end{eqnarray}
where $\xi$ is a parameter. The optimal strategy under this
scenario is the following.
\begin{enumerate}
\item At the first time slot, choose channel $1$, if $\xi>1/2$. If
$\xi=1/2$, randomly choose channel $1$ or channel $2$. Otherwise
choose channel $2$.

\item At time slots $j\geq 2$, update the pdf based on
$\Psi(j)=\{s_1,z_{s_1},\cdots,s_{j-1},z_{s_{j-1}}\}$
using~\eqref{eq:up1} and~\eqref{eq:up2}. It is easy to see that
$f^j$ has the following form
\begin{eqnarray}
f^j(\theta_1,\theta_2)=\xi_j\delta(\theta_a,\theta_b)+(1-\xi_j)\delta(\theta_b,\theta_a).\no
\end{eqnarray}
Then, choose channel $1$ if $\xi_j>1/2$, randomly choose channel
$1$ or $2$ if $\xi_j=1/2$ and choose channel $2$ otherwise.
\end{enumerate}
The optimality of this myopic strategy was proved
in~\cite{Feldman:AnMS:62}.
\end{exmpl}

The previous myopic strategy is also optimal for some other
special scenarios. For example, if the prior pdf is
$f(\thetav)=\xi\delta(a,b)+(1-\xi)\delta(c,d)$, then any of the
following conditions ensures the optimality of the myopic
strategy~\cite{Kelley:AnS:74}: 1) $a+b=c+d=1$, 2) $a\leq b$ and
$c\leq d$, 3) $a\geq b$ and $c\geq d$.
\qed

\begin{exmpl}\label{exmpl:2}(One Known Channel)
We have $N=2$ channels with independent traffic distributions.
Channel 1 and channel 2 are independent. Moreover, $\theta_2$ is
known. The traffic pattern of channel $1$ is unknown, and the
probability density function of $\theta_1$ is given by
$f_1(\theta_1)$.

Since channel $2$ is known and is independent of channel $1$,
sensing channel $2$ will not provide the cognitive user with any
new information. Hence, once the cognitive user starts accessing
channel $2$ (meaning that at a certain stage, sensing channel $2$
is optimal), there would be no reason to return to channel $1$ in
the optimal strategy. A generalized version of this assertion was
first proved in Lemma 4.1 of~\cite{Bradt:AnMS:56}. Restated in our
channel selection setup, we have the following lemma.
\begin{lem}\label{lem:rule}
In the optimal medium access strategy, once the cognitive user
starts accessing channel $2$, it should keep picking the same
channel in the remaining time slots, regardless of the outcome of
the sensing process.\qed
\end{lem}

This lemma essentially converts the channel selection problem to
an optimal stopping problem~\cite{Chow:Book:71,Ferguson}, where we
only need to focus on the strategies that decide at which
time-slot we should stop sensing channel $1$, if it is ever
accessed. The following lemma derives the optimal stopping rule.

\begin{lem}\label{lem:thre}
For any $f_1(\theta_1)$ and any $T$, if $\theta_2\geq \Lambda(f_1,
T)$, then we should sense channel 2. Here
\begin{eqnarray}\label{eq:lambda}
\Lambda(f_1,T)=\max\limits_{\Gamma(f_1)=1}\frac{\mathbb{E}_{f_1}\left\{\sum_{j=1}^M
Z_{1}(j)\right\}}{\mathbb{E}_{f_1}\{M\}},
\end{eqnarray}
where $\Gamma$ are the set of strategies that start with channel
$1$ and never switch back to channel $1$ after selecting channel
$2$; and $M$ is a random number that represents the last time slot
in which channel $1$ is sensed, when the cognitive user follows a
strategy in $\Gamma$.
\end{lem}
\begin{proof}
This result follows as a direct application of Theorem 5.3.1 and
Corollary 5.3.2 of~\cite{Berry:Book:85}.
\end{proof}
One can now combine Lemma~\ref{lem:rule} and Lemma~\ref{lem:thre}
to obtain the following optimal strategy.

\begin{enumerate}

\item At any time slot $j$, if channel $2$ was sensed at time slot
$j-1$, keep sensing channel 2.

\item If channel $1$ was sensed at time slot $j-1$, update the
pdf $f^j$ using~\eqref{eq:up1} and~\eqref{eq:up2} and
compute $\Lambda(f_1^j, T-j+1)$ using~\eqref{eq:lambda}. If
$\Lambda(f_1^j,T-j+1)<\theta_2$, switch to channel $2$; otherwise,
keep sensing channel $1$.\qed
\end{enumerate}
\end{exmpl}

\begin{exmpl} (Independent Channels)\label{exmpl:ind}

We have $N$ independent channels with
$f(\thetav)=\prod\limits_{i=1}^Nf_i(\theta_i)$. This case has a
simple form of solution in the asymptotic scenario $T\rightarrow
\infty$ assuming the following discounted form for the utility
function
\begin{eqnarray}
W=\mathbb{E}_f\left\{\sum\limits_{j=1}^{\infty}\alpha^{j}BZ_{S(j)}(j)\right\},\no
\end{eqnarray}
where $0<\alpha<1$ is a discount factor. As discussed in the
introduction, this scenario has been considered
in~\cite{Motamedi:ETTRT:07}, and the optimal strategy for this
scenario is the following.
\begin{enumerate}

\item If channel $l$ was selected at time slot $j-1$, then we get
the updated pdf $f_l^j$ using equations~\eqref{eq:up1}
and~\eqref{eq:up2}, based on the sensing result $z_l(j-1)$. For
other channels, we let $f_i^j=f_i^{j-1},\forall i\neq l,
i\in\mathcal{N}$. That is we only update the pdf of the
channel which was just accessed (due to the independence
assumption).

\item For each channel, we calculate an index using the following
equation
\begin{eqnarray}
\Lambda_i(f_i^j)=\max\limits_{\Gamma(f_i^j)=i}\frac{\mathbb{E}_{f_i^j}\left\{\sum_{j=1}^M
\alpha^{j}Z_{1}(j)\right\}}{\mathbb{E}_{f_i^j}\{\sum_{j=1}^M
\alpha^{j}\}},\no
\end{eqnarray}
where $\Gamma$ is the set of strategies for the equivalent
One-Known-Channel selection problem (with channel $i$ having the
unknown parameter) and $M$ is a random number corresponding to the
last time slot in which channel $i$ will be selected in the
equivalent One-Known-Channel case. $\Lambda_i$ is typically
referred to as the Gittins Index~\cite{Gittins:Book:74}.

\item Choose the channel with the largest Gittins index to sense
at time slot $j$.
\end{enumerate}

The optimality of this strategy is a direct application of the
elegant result of Gittins and Jones~\cite{Gittins:Book:74}.
Computational methods for evaluating the Gittins Index $\Lambda$
could be found in~\cite{Katehakis:MOR:87} and references therein.
\end{exmpl}
\subsection{Non-parametric Asymptotic Analysis and Asymptotically Optimal Strategies}\label{sec:suboptimal}
The optimal solution developed in Section~\ref{sec:opgen} suffers
from a prohibitive computational complexity. In particular, the
dimensionality of our search dimension grows exponentially with
the block length $T$. Moreover, one can envision many practical
scenarios in which it would be difficult for the cognitive user to
obtain the prior information $f(\thetav)$. This motivates our
pursuit of low complexity non-parametric protocols which maintain
certain optimality properties. Towards this end, we study in the
following the asymptotic performance of several low complexity
approaches. In this section, we analyze non-parametric schemes that
do not explicitly use $f(\thetav)$, thus the rules $\Gamma$
considered in this section depend only on $\Psi(j)$ explicitly. We
aim to develop schemes that have low complexity but still maintain
certain optimality. Towards this end, we study the asymptotic
performance of schemes as the block length $T$ increases. This
section will be concluded with our asymptotically optimal
non-parametric protocols which require only linear computational
complexity.

For a certain strategy $\Gamma$, the expected number of bits the
cognitive user is able to transmit through a block with certain
parameters $\thetav$ is
\begin{eqnarray}
\mathbb{E}\left\{\sum \limits_{j=1}^{T}BZ_{S(j)}(j)\right\}
=\sum\limits_{j=1}^{T}B\sum\limits_{i=1}^N\theta_i
\text{Pr}\left\{\Gamma(\Psi(j))=i\right\}.\no
\end{eqnarray}
Recall that $\Gamma(\Psi(j))=i$ means that, following strategy
$\Gamma$, the cognitive user should choose channel $i$ at time
slot $j$, based on the available information $\Psi(j)$. Here
$\text{Pr}\left\{\Gamma(\Psi(j))=i\right\}$ is the probability
that the cognitive user will choose channel $i$ at time slot $j$,
following the strategy $\Gamma$.

Compared with the idealistic case where the exact value of
$\thetav$ is known, in which the optimal strategy for the
cognitive user is to always choose the channel with the largest
free probability, the loss entailed by $\Gamma$ is given by
\begin{eqnarray}
L(\thetav;\Gamma)=\sum\limits_{j=1}^TB\theta_{i^*}-\sum\limits_{j=1}^{T}B\sum\limits_{i=1}^N\theta_i
\text{Pr}\left\{\Gamma(\Psi(j))=i\right\},\no
\end{eqnarray}
where $\theta_{i^*}=\max\{\theta_1,\cdots,\theta_N\}$. We say that
a strategy $\Gamma$ is consistent, if for any $\thetav\in[0,1]^N$,
there exists $\beta<1$ such that $L(\thetav;\Gamma)$ scales
as\footnote{In this paper, we use Knuth's asymptotic notations
1)$g_1(N)=o(g_2(N))$ means $\forall c>0,\exists N_{0}, \forall
N>N_{0}, g_1(N)<c g_2(N) $, 2) $g_1(N)=\omega(g_2(N))$ means
$\forall c>0,\exists N_{0}, \forall N>N_{0}, g_2(N)< c g_1(N) $,
3) $g_1(n)=O(g_2(N))$ means $\exists c_{2}\geq c_{1}>0, N_{0}$,
$\forall N>N_{0}, c_{1}g_2(N)\leq g_1(N)\leq c_{2}g_2(N)$.}
$O(T^{\beta})$. For example, consider a royal scheme in which the
cognitive user selects channel $i$ at the beginning of a block and
sticks to it. If $\theta_i$ is the largest one among $\thetav$,
$L(\thetav;\Gamma)=0$. On the other hand, if $\theta_i$ is not the
largest one, $L(\thetav;\Gamma)\sim O(T)$. Hence, this royal
scheme is not consistent. The following lemma characterizes the
fundamental limits of any consistent scheme.

\begin{lem}\label{lem:lowerbound} For any $\thetav$ and any consistent strategy $\Gamma$, we have
\begin{eqnarray}
\lim\inf\limits_{T\rightarrow\infty} \frac{L(\thetav;\Gamma)}{\ln
T}\geq B\sum \limits_{i\in\mathcal{N}\backslash
\{i^*\}}\frac{\theta_{i^*}-\theta_i}{D(\theta_i||\theta_i^*)},
\end{eqnarray}
where $D(\theta_i||\theta_l)$ is the Kullback-Leibler divergence
between the two Bernoulli random variables with parameters
$\theta_i$ and $\theta_l$ respectively:
$$D(\theta_i||\theta_l)=\theta_i\ln\left(\frac{\theta_i}{\theta_l}\right)+(1-\theta_i)\ln\left(\frac{1-\theta_i}{1-\theta_l}\right).$$
\end{lem}
\begin{proof}
The proof is an application of a theorem proved
in~\cite{Lai:AAM:85}. More specifically, for a general bandit
problem, let $x$ be the random payoff obtained by choosing bandit
$i$ (not necessarily Bernoulli), and we also let $h_{\theta_i}(x)$ be
the pdf of $x$ for a given $\theta_i$.

Let $\mu_i$ denote the average payoff of bandit $i$, i.e.
$$\mu_i=\int xh_{\theta_i}(x)dx,$$
and note that the Kullback-Leibler divergence between bandit $i$
and $l$ is given by
\begin{eqnarray}
D(\theta_i||\theta_l)=\int\Big[\ln h_{\theta_i}(x)-\ln
h_{\theta_l}(x)\Big]h_{\theta_i}(x)d x.\no
\end{eqnarray}

Let $i^*=\arg\max\limits_{i\in\mathcal{N}}\mu_i$, i.e., the index
of the channel with the largest average payoff. Under mild
regularity conditions on $h_{\theta_i}(x)$, it has been proved in
Theorem 1 of~\cite{Lai:AAM:85} that for any consistent strategy
$\Gamma$
\begin{eqnarray}\label{eq:lowerbound}
\lim\inf\limits_{T\rightarrow\infty} \frac{L(\thetav;\Gamma)}{\ln
T}\geq \sum \limits_{i\in\mathcal{N}\backslash
\{i^*\}}\frac{\mu_{i^*}-\mu_i}{D(\theta_i||\theta_i^*)}.
\end{eqnarray}
In our cognitive radio channel selection problem, given $\thetav$,
$x$ is a random variable with
$$h_{\theta_i}(x)=\theta_i\delta(B)+(1-\theta_i)\delta(0);$$
hence $\mu_i=B\theta_i$, and
$$D(\theta_i||\theta_l)=\theta_i\ln\left(\frac{\theta_i}{\theta_l}\right)+(1-\theta_i)\ln\left(\frac{1-\theta_i}{1-\theta_l}\right).$$
Substituting these parameters into~\eqref{eq:lowerbound}, the
proof is complete.
\end{proof}

Lemma~\ref{lem:lowerbound} shows that the loss of any consistent
strategy scales at least as $\omega(\ln T)$. An intuitive
explanation of this loss is that we need to spend at least $O(\ln
T)$ time slots on sampling each of the channels with smaller
$\theta_i$, in order to get a reasonably accurate estimate of
$\thetav$, and hence, use it to determine the channel having the
largest $\theta_i$ to sense. We say that a strategy $\Gamma$ is
order optimal if $L(\thetav;\Gamma)\sim O(\ln T)$.

Now, the first question that arises is whether there exists order
optimal strategies. As shown later in this section, we can design
suboptimal strategies that have loss of order $O(\ln T)$. Thus the
answer to this question is affirmative. Before proceeding to the
proposed low complexity order-optimal strategy, we first analyze
the loss order of some heuristic strategies which may appear
appealing in certain applications.

The first simple rule is the random strategy $\Gamma_r$ where, at
each time slot, the cognitive user randomly chooses a channel from
the available $N$ channels. The fraction of time slots the
cognitive user spends on each channel is therefore $1/N$, leading
to the loss
$$L(\thetav;\Gamma_r)=\frac{B\sum\limits_{i=1}^N (\theta_{i^*}-\theta_i)}{N}T\sim O(T).$$

The second one is the myopic rule $\Gamma_g$ in which the
cognitive user keeps updating $f^j(\thetav)$, and chooses the
channel with the largest value of
$$\hat{\theta}_i=\int\theta_i f^j(\thetav)d\thetav$$
at each stage. Since there are no converge guarantees for the
myopic rule, that is $\hat{\thetav}$ may never converge to
$\thetav$ due to the lack of sufficiently many samples for each
channel~\cite{Kumar:SJCO:85}, the loss of this myopic strategy is
$O(T)$.

The third protocol we consider is {\em staying with the winner and
switching from the loser rule} $\Gamma_{SW}$ where the cognitive
user randomly chooses a channel in the first time slot. In the
succeeding  time-slots 1) if the accessed channel was found to be
free, it will choose the same channel to sense; 2) otherwise, it
will choose one of the remaining channels based on a certain
switching rule.

\begin{lem}\label{lem:sw}
No matter what the switching rule is, $L(\thetav;\Gamma_{SW})\sim
O(T)$.
\end{lem}
\begin{proof}
Let $i^*=\arg\max\limits_{i\in\mathcal{N}}\theta_i$ and
$i^{**}=\arg\max\limits_{i\in\mathcal{N}\backslash\{i^*\}}\theta_i$,
i.e., $i^*$ is the best channel, and $i^{**}$ is the second best
channel. To avoid trivial conditions, without loss of generality
we assume that $\theta_{i^*}\neq\theta_{i^{**}}$ and
$\theta_{i^*}\neq 1$. We can upper bound the performance of the
staying with the winner and switching from the loser rule by
assuming that the cognitive user has the following extra
knowledge.
\begin{enumerate}

\item In the first time slot, the cognitive user is able to choose
$i^*$ correctly.

\item Once $i^*$ is sensed busy, the cognitive user somehow knows which
channel is the second best, and switches to $i^{**}$.

\item Once $i^{**}$ is sensed busy, the cognitive user is always able to
switch back to $i^*$.
\end{enumerate}
We denote this optimistic rule by $\Gamma_{SW}^*$. With any
realistic switching rule $\Gamma_{SW}$, we have
$$L(\thetav;\Gamma_{SW})\geq L(\thetav; \Gamma_{SW}^*).$$

Now with the optimistic rule $\Gamma_{SW}^*$, the system can be
modelled as the following Markov process as shown in
Figure~\ref{fig:markov}, in which we have two states: 1) sensing
channel $i^*$ and 2) sensing channel $i^{**}$. The transition
probability matrix is
\begin{eqnarray}
P=\left[\begin{array}{ll} \theta_{i^*}, &
1-\theta_{i^*}\\1-\theta_{i^{**}}, &
\theta_{i^{**}}\end{array}\right].\no
\end{eqnarray}
\begin{figure}[thb]
\centering
\includegraphics[width=0.45 \textwidth]{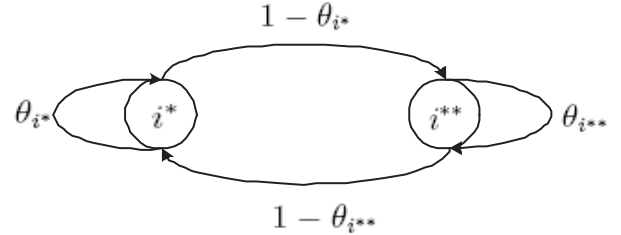}
\caption{A Markov process representation of the optimistic
strategy $\Gamma_{SW}^*$.} \label{fig:markov}
\end{figure}
The probability $P_{i^{**}}$ that the cognitive user will sense
channel $i^{**}$ can be obtained by the solving the following
stationary equation
\begin{eqnarray}
P_{i^{**}}=(1-\theta_{i^*})(1-P_{i^{**}})+\theta_{i^{**}}P_{i^{**}},\no
\end{eqnarray}
from which we obtain
\begin{eqnarray}
P_{i^{**}}=\frac{1-\theta_{i^*}}{1-\theta_{i^*}+1-\theta_{i^{**}}}.\no
\end{eqnarray}
Hence in the nontrivial cases, we have
\begin{eqnarray}
L(\thetav;\Gamma_{SW}^*)=BP_{i^{**}}(\theta_{i^*}-\theta_{i^{**}})T,\no
\end{eqnarray}
implying that, for any switching rule, $L(\thetav;\Gamma_{SW})\sim
O(T)$.
\end{proof}

There are several strategies that have loss of order $O(\ln T)$.
We adopt the following linear complexity strategy which was
proposed and analyzed in~\cite{Auer:ML:02}.

\begin{rl}(Order optimal  single index strategy)

The cognitive user maintains two vectors $\mathbf{X}$ and
$\mathbf{Y}$, where each $X_i$ records the number of time slots
for which the cognitive user has sensed channel $i$ to be free,
and each $Y_i$ records the number of time slots for which the
cognitive user has chosen channel $i$ to sense. The strategy works
as follows.

\begin{enumerate}

\item Initialization: at the beginning of each block, sense each
channel once.

\item After the initialization period, the cognitive user obtains
an estimation $\hat{\thetav}$ at the beginning of time slot $j$,
given by
\begin{eqnarray}
\hat{\theta}_i(j)=\frac{X_i(j)}{Y_i(j)},\no
\end{eqnarray}
and assigns an index
$$\Lambda_i(j)=\hat{\theta}_i(j)+\sqrt{\frac{2\ln j}{Y_i(j)}}$$
to the $i^{th}$ channel. The cognitive user chooses the channel
with the largest value of $\Lambda_i(j)$ to sense at time slot
$j$. After each sensing, the cognitive user updates $\mathbf{X}$
and $\mathbf{Y}$.

\end{enumerate}

The intuition behind this strategy is that as long as $Y_i$ grows
as fast as $O(\ln T)$, $\Lambda_i$ converges to the true value of
$\theta_i$ in probability, and the cognitive user will choose the
channel with the largest $\theta_i$ eventually. The loss of $O(\ln
T)$ comes from the time spent on sampling the inferior channels in
order to learn the value of $\thetav$. This price, however, is
inevitable as established in the lower bound of
Lemma~\ref{lem:lowerbound}.\qed
\end{rl}

Finally, we observe that the difference between the myopic rule
and the order optimal single index rule is the additional term
$\sqrt{2\ln j/Y_i(j)}$ added to the current estimate
$\hat{\theta}_i$. Roughly speaking, this additional term
guarantees enough sampling time for each channel, since if we
sample channel $i$ too sparsely, $Y_i(j)$ will be small, which
will increase the probability that $\Lambda_i$ is the largest
index. When $Y_i(j)$ scales as $\ln T$, $\hat{\theta}_i$ will be
the dominant term in the index $\Lambda_i$, and hence the channel
with the largest $\theta_i$ will be chosen much more frequently.

\section{Multi User--Single Channel}\label{sec:mulp}
The presence of multiple cognitive users adds an element of
competition to the problem. In order for a cognitive user to get
hold of a channel now, it must be free from the primary traffic
and the other competing cognitive users. More rigorously, we
assume the presence of a set $\mathcal{K}=\{1,\cdots,K\}$ of
cognitive users and consider the distributed medium access
decision processes at the multiple users with no prior
coordination. We denote $\mathcal{K}_i(j)\subseteq\mathcal{K}$ as
the random set of users who choose to sense channel $i$ at time
slot $j$. We assume that the users follow a generalized version of
the Carrier Sense Multiple Access/Collision Avoidance (CSMA-CA)
protocol to access the channel after sensing the main channel to
be free, i.e., if channel $i$ is free, each user $k$ in the set
$\mathcal{K}_i(j)$ will generate a random number $t_k(j)$
according to a certain probability density function $g$, and wait
the time specified by the generated random number. At the end of
the waiting period, user $k$ senses the channel again, and if it
is found free, the packet from user $k$ will be transmitted. The
probability that user $k$ in the set $\mathcal{K}_i(j)$ gains
access to the channel is the same as the probability that $t_k(j)$
is the smallest random number generated by the users in the set
$\mathcal{K}_i(j)$. Thus, the throughput user $k$ achieves in a
block is
\begin{eqnarray}
W_k=\sum\limits_{j=1}^{T}BZ_{S_k(j)}(j)I\left\{k=\arg\min\limits_{q\in\mathcal{K}_{S_k(j)}(j)}
t_q(j)\right\}.\no
\end{eqnarray}

Therefore, user $k$ should devise sensing rule $\Gamma_k$ that
maximizes
\begin{eqnarray}
\mathbb{E}\left\{W_k\right\}=\mathbb{E}\left\{\sum\limits_{j=1}^{T}BZ_{S_k(j)}(j)I\left\{k=\arg\min\limits_{q\in\mathcal{K}_{S_k(j)}(j)}
t_q(j)\right\}\right\}.\no
\end{eqnarray}

Clearly, with multiple cognitive users, it is not optimal anymore
for all the users to always choose the channel with the largest
$\theta_i$ to sense. In particular, if all the users choose the
channel with the largest $\theta_i$, the probability that a given
user gains control of the channel decreases, while potential
opportunities in the other channels in the primary network are
wasted.

\subsection{Known $\thetav$ Case}\label{known}
To enable a succinct presentation, we first consider the case in
which the values of $\thetav$ are known to all the cognitive
users. The users distributively choose channels to sense and
compete for access if the channels are free.

\subsubsection{The Optimal Symmetric Strategy}

Without loss of generality, we consider a mixed strategy where
user $k$ will choose channel $i$ with probability $p_{k,i}$.
Furthermore, we let $\mathbf{p}_k=[p_{k,1},\cdots,p_{k,N}]$ and
consider the symmetric solution in which
$\mathbf{p}=\mathbf{p}_1=\cdots=\mathbf{p}_K$. The symmetry
assumption implies that all the users in the network
distributively follow the same rule to access the spectral
opportunities present in the primary network, in order to maximize
the same average throughput each user can obtain. The following
result derives the optimal solution in this situation.

\begin{lem}\label{lem:poweroptimal}
For a cognitive network with $K>1$ cognitive users and $N$
channels with probability $\thetav$ of being free, the optimal
$\mathbf{p}^*$ is given by
\begin{eqnarray}
p_i^*=\left\{\begin{array}{ll}
\left\{1-\left(\frac{\lambda^*}{K\theta_i}\right)^{1/(K-1)}\right\}^{+},&\quad
\text{for}\quad \theta_i>0,\\
0,& \quad \text{for}\quad \theta_i=0,\end{array}\right.\no
\end{eqnarray}
where $\lambda^*$ is a constant such that $\sum\limits_{i=1}^{N}
p_i^*=1$. Here $\{x\}^+=\max\{0,x\}$.
\end{lem}
\begin{proof}
With a strategy $\mathbf{p}$, the probability that user $k$
chooses channel $i$ and, at the same time, there are $l$ other
users choosing channel $i$ to sense is
\begin{eqnarray}
p_i {{K-1}\choose l} p_i^l(1-p_i)^{K-1-l}\no.
\end{eqnarray}
Under this scenario, the average bits transmitted at one slot of each user is
$B\theta_i/(l+1)$,
Hence, the average throughput $W_k$ of user $k$ is
\begin{eqnarray}
W_k=T\sum\limits_{i=1}^N\frac{B\theta_i}{l+1}\sum\limits_{l=0}^{K-1}p_i{{K-1}\choose
l}p_i^l(1-p_i)^{K-1-l}.\no
\end{eqnarray}
Based on our symmetry assumption, we drop the subscript $k$ and
write the average throughput of each user as $W$ leading to
\begin{eqnarray}
W&=&BT\sum\limits_{i=1}^Np_i\theta_i\sum\limits_{l=0}^{K-1}{{K-1}\choose
l} \frac{p_i^l(1-p_i)^{K-1-l}}{l+1}\no\\
&=&BT\sum\limits_{i=1}^Np_i\theta_i\sum\limits_{l=0}^{K-1}\frac{(K-1)!}{l!(K-1-l)!}\frac{p_i^l(1-p_i)^{K-1-l}}{l+1}\no\\
&=&BT\sum\limits_{i=1}^N\frac{\theta_i}{K}\sum\limits_{l=0}^{K-1}{K\choose {l+1}}p_i^{l+1}(1-p_i)^{K-1-l}\no\\
&=&BT\sum\limits_{i=1}^N\frac{\theta_i}{K}\left\{\sum\limits_{l^{'}=0}^{K}{K\choose l^{'}}p_i^{l^{'}}(1-p_i)^{K-l^{'}}-(1-p_i)^{K}\right\}\no\\
&=&BT\sum\limits_{i=1}^N\frac{\theta_i}{K}\left\{1-(1-p_i)^{K}\right\}\no.
\end{eqnarray}
Now, we should solve the following optimization problem
\begin{eqnarray}
\max\quad &&
W=BT\sum\limits_{i=1}^N\frac{\theta_i}{K}\left\{1-(1-p_i)^{K}\right\},\no\\
\text{s.t.}\quad&& \sum\limits_{i=1}^N p_i=1,\no\\
&& \mathbf{p}\geq \mathbf{0}.\no
\end{eqnarray}
This optimization problem is equivalent to the following:
\begin{eqnarray}\label{eq:op}
\min \quad && y=\sum\limits_{i=1}^N\theta_i(1-p_i)^{K},\no\\
\text{s.t.}\quad&& \sum\limits_{i=1}^N p_i=1,\\
&& \mathbf{p}\geq \mathbf{0}.\no
\end{eqnarray}
Since
\begin{eqnarray}
\frac{\partial^2 y}{\partial^2
p_i}=\theta_iK(K-1)(1-p_i)^{K-2}\geq0,\no
\end{eqnarray}
for $0\leq p_i\leq 1$, $y$ is a convex function of $\mathbf{p}$ in
the region of interest, i.e. $\mathbf{p}\in [0,1]^N$. Also, the
constraints are the intersection of a convex set and a linear
constraint. Therefore, our problem reduces to a convex
optimization problem whose Karush-Kuhn-Tucker (KKT)
conditions\cite{Boyd:Book:04} for optimality are
\begin{eqnarray}
\mathbf{p}^*&\geq& \mathbf{0},\no\\
\sum \limits_{i=1}^Np^*_i&=&1,\no\\
p_i^*\left(\lambda^*-K\theta_i(1-p_i^*)^{K-1}\right)&=&0,\no\\
\lambda^*&\geq& K\theta_i(1-p_i^*)^{K-1},\no
\end{eqnarray}
where $\lambda^*$ is the Lagrange multiplier.

It is easy to check that if $K>1$,
\begin{eqnarray}\label{eq:opsolution}
p_i^*=\left\{\begin{array}{ll}
\left\{1-\left(\frac{\lambda^*}{K\theta_i}\right)^{1/(K-1)}\right\}^{+}&\quad
\text{for}\quad \theta_i>0,\\
0& \quad \text{for}\quad \theta_i=0,\end{array}\right.
\end{eqnarray}
satisfies the KKT conditions, in which $\lambda^*$ is the constant
that satisfies $\sum p_i^*=1$.
\end{proof}

If $K=1$, then $p_{i^{*}}^*=1$, where
$i^*=\arg\max\limits_{i\in\mathcal{N}} \theta_i$, $p_l^*=0$, and $
l\in\mathcal{N}\backslash \{i^*\}$, satisfies the KKT conditions.

So, the total throughput of the $K$ cognitive users is
\begin{eqnarray}
KW&=&BKT\sum\limits_{i=1}^N\frac{\theta_i}{K}\left\{1-(1-p_i^*)^{K}\right\}\no\\
&=&BT\sum\limits_{i=1}^N\theta_i\left\{1-(1-p_i^*)^{K}\right\}.\no
\end{eqnarray}
On the other hand, the average total spectral opportunities of the
primary network is $BT\sum\limits_{i=1}^N\theta_i$. This upper
bound can be achieved by a centralized channel allocation strategy
when $K>N$ (simply by assigning one cognitive user to each
channel). Therefore, the loss of the distributed protocol as
compared with the centralized scheduling is
\begin{eqnarray}
L=BT\sum\limits_{i=1}^N\theta_i(1-p_i^*)^{K},\no
\end{eqnarray}
which is same as~\eqref{eq:op} up to a constant factor. There is
an intuitive explanation of this loss. If there is a spectral
opportunity in channel $i$ but there are no users choosing channel
$i$ to sense, a loss occurs. The probability that there is no user
choosing channel $i$ to sense is $(1-p_i^*)^K$, and hence the
probability of loss occurring at channel $i$ is
$\theta_i(1-p_i^*)^K$. To obtain further insights on the
performance of the cognitive network, we study the following
special cases.
\begin{enumerate}

\item $N\geq 1, K=1$. As stated in the above, $p_{i^*}^*=1$, and
$p_l^*=0,l\in\mathcal{N}\backslash \{i^*\}$. Hence, the user
should choose the channel with the largest free probability to
sense. And
$$L=BT\sum\limits_{i\in\mathcal{N}\backslash\{i^*\}}\theta_i.$$

\item $N=2, K=2$. Substituting $N=2$ and $K=2$
into~\eqref{eq:opsolution}, we obtain
\begin{eqnarray}
p_1^*=\theta_1/(\theta_1+\theta_2)\quad\text{and}\quad
p_2^*=\theta_2/(\theta_1+\theta_2).\no
\end{eqnarray}
Furthermore,
\begin{eqnarray}
W&=&\frac{BT\theta_1}{2}\left[1-\frac{\theta_2^2}{(\theta_1+\theta_2)^2}\right]\no\\ &&+\frac{BT\theta_2}{2}\left[1-\frac{\theta_1^2}{(\theta_1+\theta_2)^2}\right],\no\\
L&=&\frac{BT\theta_1\theta_2}{2(\theta_1+\theta_2)}.\no
\end{eqnarray}

\item $N$ is fixed, and $K\rightarrow\infty$. We have the
following asymptotic characterization.
\end{enumerate}
\begin{lem}
Let $2\leq Q\leq N$ be the number of channels for which $\theta_i>0$. We
have $p_i^*\rightarrow 1/Q$, and $L\rightarrow 0$ exponentially as
$K$ increases, i.e.,
$$L\sim O(e^{-c_1K}),$$ where $c_1=\ln \frac{Q}{Q-1}$.
\end{lem}
\begin{proof}
Without loss of generality, we assume that $\theta_i\neq 0$, for
$1\leq i\leq Q$. At the moment, we assume that (we will show that
this is true, if $K$ is large enough) if $\theta_i\neq0$
\begin{eqnarray}\label{eq:constraints}
p_i^*=\left\{1-\left(\frac{\lambda^*}{K\theta_i}\right)^{1/(K-1)}\right\}^{+}=1-\left(\frac{\lambda^*}{K\theta_i}\right)^{1/(K-1)}.\no
\end{eqnarray}
Together with $\sum\limits_{i=1}^N p_i^*=\sum\limits_{i=1}^Q
p_i^*=1$, we have
\begin{eqnarray}
(\lambda^*)^{1/(K-1)}=\frac{K^{1/(K-1)}(Q-1)}{\sum\limits_{i=1}^Q\theta_i^{-1/(K-1)}}\no
\end{eqnarray}
and
\begin{eqnarray}
p_i^*=1-\frac{(Q-1)\theta_i^{-1/(K-1)}}{\sum\limits_{i=1}^Q\theta_i^{-1/(K-1)}},\quad\text{for}\quad
1\leq i\leq Q.\no
\end{eqnarray}
To satisfy the condition $\mathbf{p}\geq \mathbf{0}$, we need to
show
\begin{eqnarray}
\frac{(Q-1)\theta_i^{-1/(K-1)}}{\sum\limits_{i=1}^Q\theta_i^{-1/(K-1)}}\leq
1,\no
\end{eqnarray}
for all $i$ with $\theta_i>0$.

With $i^*=\arg\max\limits_{i\in\mathcal{N}} \theta_i$ and
$l^*=\arg\min\limits_{1\leq l\leq Q}\theta_l$, we have for all $i$
\begin{eqnarray}
\frac{(Q-1)\theta_i^{-1/(K-1)}}{\sum\limits_{i=1}^Q\theta_i^{-1/(K-1)}}\leq
\frac{(Q-1)\theta_{l^*}^{-1/(K-1)}}{Q\theta_{i^*}^{-1/(K-1)}}.\no
\end{eqnarray}
For any $\vartheta\leq Q/(Q-1)$, if $K$ is large enough, we have
\begin{eqnarray}
\left(\frac{\theta_{i^*}}{\theta_{l^*}}\right)^{\frac{1}{K-1}}\leq
\vartheta\no
\end{eqnarray}
since
$$\lim\limits_{K\rightarrow\infty}\left(\frac{\theta_{i^*}}{\theta_{l^*}}\right)^{\frac{1}{K-1}}=1.$$
Hence, for all $1\leq i\leq Q$, we have
\begin{eqnarray}
\frac{(Q-1)\theta_i^{-1/(K-1)}}{\sum\limits_{i=1}^Q\theta_i^{-1/(K-1)}}\leq
\frac{Q-1}{Q}\vartheta\leq 1.\no
\end{eqnarray}
Now, straightforward limit calculation shows that
\begin{eqnarray}
p^*\rightarrow 1/Q,\no
\end{eqnarray}
as $K$ increases. And
\begin{eqnarray}
\lim\limits_{K\rightarrow\infty}\frac{L}{\exp^{-c_1K}}=\lim\limits_{K\rightarrow\infty}\frac{BT\sum\limits_{i=1}^Q\theta_i(1-p_i^*)^{K}}{\exp^{-c_1K}}=BT\sum\limits_{i=1}^Q\theta_i\no
\end{eqnarray}
with $c_1=\ln\frac{Q}{Q-1}$.
\end{proof}
The reason for the exponential decrease in the loss is that, as
the number of cognitive users increases, the probability that
there is no user sensing any particular channel decreases
exponentially. If $Q=1$, there is no loss of performance, since the all the user will always sense the channel with non-zero availability probability.

\subsubsection{The Game Theoretic Model}
The optimality of the distributed protocol proposed in the
previous section hinges on the assumption that all the users will
follow the symmetric rule. However, it is straightforward to see
that if a single cognitive user deviates from the rule specified
in Lemma~\ref{lem:poweroptimal}, it will be able to transmit more
bits. If this selfish perspective propagates through the network,
it may lead to a significant reduction in the overall throughput.
This observation motivates our next step in which the channel
selection problem is modeled as a non-cooperative game, where the
cognitive users are the players, the $\Gamma_k$s are the
strategies and the average throughput of each user is the payoff.
The following result derives a sufficient condition for the Nash
equilibrium~\cite{Fudenberg:Book:91} in the asymptotic scenario
$K\rightarrow \infty$.

\begin{lem}\label{lem:nash}
$(\Gamma_1,\cdots,\Gamma_K)$ is a Nash-equilibrium, if $K$ is
large and at each time slot, there are $\tau_iK$ users sensing
channel $i$, where $\tau_i$ satisfies
\begin{eqnarray}\label{eq:nash}
\tau_i=\frac{\theta_i}{\sum\limits_{i=1}^N\theta_i}.
\end{eqnarray}
At this equilibrium, each user has probability
$\frac{\sum\limits_{i=1}^N\theta_i}{K}$ of transmitting at each
time slot.
\end{lem}
\begin{proof}
We prove this by backward induction. At the last time slot $T$, if
$\tau_i$s satisfy equation~\eqref{eq:nash}, the probability of
user $k$ gaining a channel is
\begin{eqnarray}
p_k=\frac{\theta_i}{\tau_iK}=\frac{\sum\limits_{i=1}^N\theta_i}{K}.\no
\end{eqnarray}

Now, if user $k$ deviates from this strategy, and chooses channel
$i^{'}$, the number of users sensing channel $i^{'}$ is
$\tau_{i^{'}}K+1$, and the probability of user $k$ gaining the
channel is
\begin{eqnarray}
p_k^{'}=\frac{\theta_{i^{'}}}{\tau_{i^{'}}K+1}<\frac{\theta_{i^{'}}}{\tau_{i^{'}}K}=p_k.\no
\end{eqnarray}
Hence the strategy that has $\tau_iK$ users sensing channel $i$ at
time slot $T$ is a Nash equilibrium. Now, we know the optimal
strategy for the last time-slot, so we can ignore this time slot.
Then time slot $T-1$ becomes the last slot, in which this strategy
is optimal. Similarly, we show that this strategy is optimal for
all other time slots.
\end{proof}

We note that in the lemma we implicitly assume that $\tau_i K$ is an integer. In practice, this is not always true. However, since $K$ is large, rounding $\tau_i K$ to the nearest integer will have minor effects. The Nash equilibrium is also optimal from a system perspective, in the sense that this strategy maximizes the total throughput of the whole network by fully utilizing the available spectral opportunities when $K$ is large (i.e., on the average, each user will be able to transmit $\frac{BT\sum\theta_i}{K}$ bits per
block, and the total throughput of the network is $BT\sum\theta_i$).

With this equilibrium result, the cognitive users can use the
following stochastic sensing strategy to approximately work on the
equilibrium point for a large but finite $K$. Let $s_{k}(j)$ be
the channel chosen by user $k$ at time slot $j$. At each time
slot, each user independently selects channel $i$ with probability
$\tau_i=\frac{\theta_i}{\sum\limits_{i\in\mathcal{N}}\theta_i}$,
i.e., $\text{Pr}\{s_{k}(j)=i\}=\tau_i$. Then at each time slot,
the number of users sensing channel $i$ will be
$\sum\limits_{k=1}^KI\{s_{k}(j)=i\}$, where the $I\{s_{k}(j)=i\}$s
are i.i.d Bernoulli random variables. Hence, the total number of
users sensing channel $i$ is a binomial random number, and the
fraction of users sensing channel $i$ converges to $\tau_i$ in
probability as $K$ increases, i.e.
\begin{eqnarray}
\tau^{'}=\frac{\sum\limits_{k=1}^KI\{s_{k}(j)=i\}}{K}\rightarrow
\tau_i\no
\end{eqnarray}
in probability. Hence, as $K$ increases, the operating point will
converge to the Nash equilibrium in probability.

For any $K$, the probability that there is no user choosing
channel $i$ to sense is $(1-\tau_i)^K$. Hence the performance loss
compared with the centralized scheme is
\begin{eqnarray}
L=BT\sum
\theta_i(1-\tau_i)^K=BT\sum\limits_{i=1}^N\theta_i\left(\frac{\sum_{l=1}^N\theta_l-\theta_i}{\sum_{l=1}^N\theta_l}\right)^K.\no
\end{eqnarray}
It is easy to check that
\begin{eqnarray}
\lim\limits_{K\rightarrow\infty}\frac{L}{\exp^{-c_2K}}=BT\theta_{l^{*}},\no
\end{eqnarray}
where $\theta_{l^{*}}=\min\{\theta_i:\theta_i>0\}$, and $c_2=\ln
\frac{\sum \theta_i}{\sum_{l=1}^N\theta_l-\theta_{l^{*}}}$. It is now
clear that the loss of the game theoretic scheme goes to zero
exponentially, though the decay rate is smaller than that of the
scheme specified in Lemma~\ref{lem:poweroptimal}. On the other
hand, compared with the scheme in Lemma~\ref{lem:poweroptimal},
the game theoretic scheme has the advantage that the individual
cognitive users do not need to know the total number of cognitive
users $K$ in the network and, more importantly, they have no
incentive to deviate unilaterally.

\subsection{Unknown $\thetav$ Case}

If $\thetav$ is unknown, the cognitive users need to estimate
$\thetav$ (in addition to resolving their competition). Combining
the results from Sections~\ref{sec:suboptimal}~and~\ref{known}, we
design the following low complexity strategy which is
asymptotically optimal.

\begin{rl}\label{rule:game}

1) Initialization: Each user $k$ maintains the following two
vectors: $\mathbf{X}_k$, which records the number of time slots in
which user $k$ has sensed each channel to be free; and
$\mathbf{Y}_k$, which records the number of time slots in which
user $k$ has sensed each channel. At the beginning of each block,
user $k$ senses each channel once and transmits through this
channel if the channel is free and it wins the competition. Also,
set $X_{k,i}=1$, regardless of the sensing result of this stage.

2) At the beginning of time slot $j$, user $k$ estimates
$\hat{\theta}_i$ as
$$\hat{\theta}_i(j)=X_{k,i}(j)/Y_{k,i}(j),$$
and chooses each channel $i\in\mathcal{N}$ with probability
\begin{eqnarray}\label{eq:normedp}
\frac{\hat{\theta}_i(j)}{\sum\limits_{i=1}^N\hat{\theta}_i(j)}.
\end{eqnarray}
After each sensing, $\mathbf{X}_k$ and $\mathbf{Y}_k$ are
updated.\qed
\end{rl}

\begin{lem}\label{lem:adaptive}
If $K$ is large, the scheme in Rule~\ref{rule:game} converges to
the Nash equilibrium specified in Lemma~\ref{lem:nash} in
probability, as $T$ increases.
\end{lem}
\begin{proof}
$X_{k,i}$ is the sum of $Y_{k,i}$ i.i.d Bernoulli random variables
with parameter $\theta_i$. We use the following form of the
Chernoff bound. Let $X$ be the sum of $n$ independent Bernoulli
random variables with parameter $\bar{\theta}$, then
\begin{eqnarray}
\text{Pr}\left\{X\leq
(1-\delta)n\bar{\theta}\right\}<\exp\left(\frac{-n\bar{\theta}\delta^2}{2}\right)\no
\end{eqnarray}
for any $\delta<1$.

At time slot $j$, if we replace $X$ with $X_{k,i}(j)$, $n$ with
$Y_{k,i}(j)$, $\bar{\theta}$ with $\theta_i$ and let $\delta=1/2$,
then we have
\begin{eqnarray}
\text{Pr}\left\{X_{k,i}(j)\leq \frac{1}{2}\theta_i
Y_{k,i}(j)\right\}<\exp \left(-Y_{k,i}(j)\theta_i/8\right).\no
\end{eqnarray}
Hence
\begin{eqnarray}\label{eq:chernoffbound}
\text{Pr}\left\{\frac{X_{k,i}(j)}{Y_{k,i}(j)}\geq
\frac{\theta_i}{2}\right\} &\geq&
1-\exp\left(-Y_{k,i}(j)\theta_i/8\right)\no\\ &\geq&
1-\exp\left(-\theta_i/8\right),
\end{eqnarray}
since after the initialization period, $Y_{k,i}(j)\geq 1$.

Note that $Y_{k,i}(T)$ is the total number of time slots that user
$k$ has sensed channel $i$ in each block with $T$ time slots. We
have
\begin{eqnarray}
\mathbb{E}\{Y_{k,i}(T)\}&=&\mathbb{E}\left\{\sum\limits_{j=1}^{T}
I\left\{S_{k}(j)=i\right\}\right\}\no\\
&=&\sum\limits_{j=1}^{T}\mathbb{E}\left\{I\left\{S_{k}(j)=i\right\}\right\}\no\\
&=&\sum\limits_{j=1}^{T}\mathbb{E}\left\{\frac{X_{k,i}(j)/Y_{k,i}(j)}{\sum\limits_{i\in\mathcal{N}}X_{k,i}(j)/Y_{k,i}(j)}\right\}\no\\
&\overset{(a)}\geq&\sum\limits_{j=1}^{T}\mathbb{E}\left\{\frac{X_{k,i}(j)/Y_{k,i}(j)}{N}\right\}\no\\
&\overset{(b)}\geq&\sum\limits_{j=1}^{T}\frac{\theta_i(1-\exp(-\theta_i/8))}{2N}\no\\
&=& T\frac{\theta_i}{2N}(1-\exp(-\theta_i/8))=c_i T,\no
\end{eqnarray}
where (a) follows from the fact that $X_{k,i}(j)/Y_{k,i}(j)\leq
1$, and (b) follows from~\eqref{eq:chernoffbound}.

The probability that $Y_{k,i}(T)\leq
(1-\delta)\mathbb{E}\{Y_{k,i}(T)\}$ can also be bounded using the
Chernoff bounds since $Y_{k,i}(T)$ is also the sum of independent
Bernoulli random variables. In particular, we have
\begin{eqnarray}
\text{Pr}\{Y_{k,i}(T)\leq (1-\delta)\mathbb{E}\{Y_{k,i}(T)\}\}\leq
\exp\left(\frac{-\delta^2\mathbb{E}\{Y_{k,i}(T)\}}{2}\right).\no
\end{eqnarray}
On letting
$$\delta=\sqrt{\frac{\ln\mathbb{E}\{Y_{k,i}(T)\}}{\mathbb{E}\{Y_{k,i}(T)\}}},$$
we have
\begin{eqnarray}
\text{Pr}\left\{Y_{k,i}(T)\leq
\mathbb{E}\{Y_{k,i}(T)\}-\ln\mathbb{E}\{Y_{k,i}(T)\}\right\}\leq\frac{1}{c_iT}.\no
\end{eqnarray}
Using the union bound, and the weak law of large numbers,
$X_{k,i}(j)/Y_{k,i}(j)$ converges to $\theta_i$ in probability as
$T$ increases (with probability larger than $1-1/T$). The scheme
becomes the same as the known $\thetav$ case, in which we know
that the operating point is approximately at the Nash equilibrium,
if $K$ is sufficiently large.
\end{proof}

The intuition behind this scheme is that, each user will sample
each channel at least $O( T)$ times, and hence as $T$ increases,
the estimate $\hat{\thetav}$ converges to $\thetav$ in probability
implying that the unknown $\thetav$ case will eventually reduce to
the case in which $\thetav$ is known to all the users. Hence, if
$K$ is sufficiently large, the operating point converges to the
Nash equilibrium in probability.

If one can assume that the users will follow the pre-specified
rule, then we can design the following strategy that converges to
the optimal operating point in probability for any $K$, as $T$
increases.

\begin{rl}

1) Initialization: Each user $k$ maintains the following two
vectors: $\mathbf{X}_k$, which records the number of time slots in
which user $k$ has sensed each channel to be free, $\mathbf{Y}_k$,
which records the number of time slots in which user $k$ has
sensed each channel. At the beginning of each block, user $k$
senses each channel once, and transmits through this channel if
the channel is free and it wins the competition. Also, set
$X_{k,i}=1$, regardless of what the sensing result at this stage.

2) At the beginning of time slot $j\leq \ln T$, user $k$ estimates
$\hat{\theta}_i$ as
$$\hat{\theta}_i(j)=X_{k,i}(j)/Y_{k,i}(j),$$
and chooses each channel $i\in\mathcal{N}$ with probability
$\hat{\theta}_i(j)/\sum\limits_{i=1}^N\hat{\theta}_i(j)$. For
$j\geq \ln T$, the $i^{th}$ channel is sensed with probability
\begin{eqnarray}\label{eq:workpoint}
\hat{p}_i^*=\left\{1-\left(\frac{\lambda^*}{\hat{\theta_i}}\right)^{1/(K-1)}\right\}^{+}.
\end{eqnarray}
After each sensing, $\mathbf{X}_k$ and $\mathbf{Y}_k$ are
updated.\qed
\end{rl}

\begin{lem}\label{lem:adaptiveone}
The proposed scheme converges in probability to the optimal
operating point specified in Lemma~\ref{lem:poweroptimal}, as $T$
increases.
\end{lem}
\begin{proof}
Following the same steps as the proof of Lemma~\ref{lem:adaptive},
one can show that after $O(\ln T)$ times slots, $\hat{\thetav}$
converges to $\thetav$ in probability as $T$ increases. Hence the
operating point specified by~\eqref{eq:workpoint} converges in
probability to the optimal point specified in
Lemma~\ref{lem:poweroptimal} as $T$ increases.
\end{proof}

\section{Multi-Channel Cognitive Users}\label{sec:multisensing}
In certain scenarios, cognitive users may be able to sense more
than one channel simultaneously. To simplify the presentation, we
assume the presence of only a single cognitive user capable of
sensing, and subsequently utilizing, $M\leq N$ channels
simultaneously. Let $\mathcal{M}(j)$ be the set of channels the
cognitive user selects to sense at time slot $j$, where
$|\mathcal{M}(j)|=M$. The average number of bits that the
cognitive user is able to send over a block is therefore
\begin{eqnarray}
\mathbb{E}\{W\}=\mathbb{E}\left\{\sum\limits_{j=1}^{T}\sum\limits_{S(j)\in\mathcal{M}(j)}BZ_{S(j)}(j)\right\}.\no
\end{eqnarray}

At the beginning of time slot $j$, the cognitive user can update
the pdf $f^j(\thetav)$ according to~\eqref{eq:up1}
and~\eqref{eq:up2}. Similar to Lemma~\ref{lem:optimal}, the
optimal solution can be characterized by the following optimality
condition
\begin{eqnarray}\label{eq:multisolution}
\hspace{-5mm}V^*(f,T)&\hspace{-2mm}=&\hspace{-2mm}\max\limits_{\mathcal{M}(1)\subseteq\mathcal{N},|\mathcal{M}(1)|=M}\mathbb{E}_{f}\Bigg\{\sum\limits_{s(1)\in\mathcal{M}(1)}BZ_{s(1)}\no\\
&&\hspace{4mm}+V^*\left(f_{\{Z_{s(1)}:s(1)\in\mathcal{M}(1)\}},T-1\right)\Bigg\}.
\end{eqnarray}
Here, $f_{\{Z_{s(1)}:s(1)\in\mathcal{M}(1)\}}$ is the updated
density after observing the sensing output of the channels
$s(1)\in \mathcal{M}(1)$. We can then follow the same procedure
described for the single-channel sensing scenario to obtain the
optimal strategy $\Gamma^*$ according to~\eqref{eq:multisolution}.
In the following, however, we focus on low complexity
non-parametric strategies that are asymptotically optimal.

If $\thetav$ is known, the cognitive user will choose the $M$
channels with the largest $\theta$'s to sense. Without loss of
generality, we assume $\theta_1\geq\theta_2\geq\cdots\geq
\theta_N$. Hence, for any strategy $\Gamma$, the loss is
\begin{eqnarray}
L(\thetav;\Gamma)=\sum\limits_{j=1}^T\sum\limits_{i=1}^MB\theta_{i}-\sum\limits_{j=1}^{T}B\sum\limits_{i=1}^N\theta_i
P\left\{i\in \mathcal{M}(j)\right\},\no
\end{eqnarray}
We have the following order-optimal simple single-index strategy.
\begin{rl}\label{rule:multisense}
The cognitive user maintains two vectors $\mathbf{X}$ and
$\mathbf{Y}$, where each $X_i$ is the number of time slots in
which the cognitive user has sensed channel $i$ to be free, and
each $Y_i$ is the number of time slots in which the cognitive user
has chosen channel $i$ to sense. The strategy works as follows.

\begin{enumerate}

\item Initialization: at the beginning of each block, each channel
is sensed once. This initialization stage takes $\lceil N/M\rceil$
time slots, in which $\lceil x\rceil$ denotes the least positive
integer that is no less than $x$.

\item After the initialization period, the cognitive user obtains
an estimation $\hat{\thetav}$ at the beginning of time slot $j$
given by
\begin{eqnarray}
\hat{\theta}_i(j)=\frac{X_i(j)}{Y_i(j)},\no
\end{eqnarray}
and assigns an index
$$\Lambda_i(j)=\hat{\theta}_i(j)+\sqrt{\frac{2\ln j}{Y_i(j)}}$$
to the $i^{th}$ channel. The cognitive user orders these
$\Lambda_i(j)$s and selects the $M$ channels with the largest
$\Lambda_i(j)$s to sense. After each sensing, the cognitive user
updates $\mathbf{X}$ and $\mathbf{Y}$.
\end{enumerate}
\end{rl}
\begin{lem}
Rule~\ref{rule:multisense} is asymptotically optimal and
$L(\thetav,\Gamma)\sim O(\ln T).$
\end{lem}
\begin{proof}
We bound $Y_i(T)$ for $i\geq M+1$, i.e., the channels that are not
among the channels having the $M$ largest values of $\theta$. Note
that $Y_i(T)$ is the total number of time slots in which the
cognitive user has sensed channel $i$ in a block with $T$ time
slots. We have
\begin{eqnarray}
Y_i(T)&=&1+\sum\limits_{j=\lceil N/M\rceil+1}^T
I\left\{i\in\mathcal{M}(j)\right\}\no\\
&\leq& m+\sum\limits_{j=\lceil N/M\rceil+m}^T
I\left\{i\in\mathcal{M}(j)\Big|Y_i(j)\geq m\right\},\no
\end{eqnarray}
for any $m\geq 1$, where $I\{x|y\}$ is the conditional indicator
function, which equals 1 if, conditioning on $y$, $x$ is
satisfied, and otherwise equals 0. Since $Y_i(j)\geq m$, it
follows that $i\in\mathcal{M}(j)$ only if $\Lambda_i(j)$ is among
the $M$ largest indices. Hence, a necessary condition for
$i\in\mathcal{M}(j)$ is
$$\Lambda_{i}(j)\geq\min\{\Lambda_{l}(j):1\leq l\leq M\}.$$
Otherwise, if $$\Lambda_{i}(j)<\min\{\Lambda_{l}(j):1\leq l\leq
M\},$$ then the indices of these $M$ channels are already larger
than that of channel $i$, and channel $i$ will not be selected.
Thus
\begin{eqnarray}
&&\hspace{-13mm}I\left\{i\in\mathcal{M}(j)\Big|Y_i(j)\geq
m\right\}\no
\\&\leq& I\left\{\Lambda_{i}(j)\geq\min\{\Lambda_{l}(j):1\leq
l\leq
M\}\Big|Y_i(j)\geq m\right\}\no\\
&\leq&\sum\limits_{l=1}^MI\left\{\Lambda_{i}(j)\geq
\Lambda_{l}(j)\Big|Y_i(j)\geq m\right\}.\no
\end{eqnarray}
Hence
\begin{eqnarray}
Y_i(T)&\hspace{-3mm}\leq&\hspace{-3mm} m+\sum\limits_{j=\lceil
N/M\rceil+m}^T\sum\limits_{l=1}^M I\left\{\Lambda_{i}(j)\geq
\Lambda_{l}(j)\Big|Y_i(j)\geq
m\right\}\no\\
&\hspace{-12mm}\leq&\hspace{-8mm}
\sum\limits_{l=1}^M\left\{m+\sum\limits_{j=\lceil N/M\rceil+m}^T
I\left\{\Lambda_{i}(j)\geq \Lambda_{l}(j)\Big|Y_i(j)\geq
m\right\}\right\}\no.
\end{eqnarray}
In order for $\Lambda_{i}(j)\geq \Lambda_{l}(j)$, one of the
following three conditions must be satisfied
\begin{eqnarray}
\Lambda_l(j)\leq \theta_l,\; \Lambda_i(j)\geq
\theta_i+2\sqrt{\frac{2\ln j}{Y_i(j)}}, \;\text{or}\;\theta_l\leq
\theta_i+2\sqrt{\frac{2\ln j}{Y_i(j)}}.\no
\end{eqnarray}
One can easily check that, if none of these three conditions is
satisfied, we will have $\Lambda_i(j)<\Lambda_l(j)$. In the
following, we bound the probability of each event.
\begin{eqnarray}\label{eq:bound}
&&\hspace{-10mm}\text{Pr}\left\{\Lambda_l(j)\leq
\theta_l\Big|Y_i(j)\geq m\right\}\no\\
&=&\text{Pr}\left\{\hat{\theta}_l+\sqrt{\frac{2\ln j}{Y_l(j)}}\leq
\theta_l\Bigg|Y_i(j)\geq
m\right\}\no\\
&\leq&\text{Pr}\left\{\Big|\hat{\theta}_l-\theta_l\Big|\geq\sqrt{\frac{2\ln
j}{Y_l(j)}}\Bigg|Y_i(j)\geq
m\right\}\no\\
&=&\sum\limits_{q=1}^{j}\text{Pr}\{Y_l(j)=q\}\no\\
&&\hspace{6mm}\text{Pr}\left\{\Big|\hat{\theta}_l-\theta_l\Big|\geq\sqrt{\frac{2\ln
j}{Y_l(j)}}\Bigg|Y_i(j)\geq
m, Y_l(j)=q\right\}\no\\
&\leq&\sum\limits_{q=1}^{j}\text{Pr}\left\{\Big|\hat{\theta}_l-\theta_l\Big|\geq\sqrt{\frac{2\ln
j}{Y_l(j)}}\Bigg|Y_l(j)=q\right\}\no\\
&\leq&2j\exp^{-4\ln j}\\ &=&2j^{-3},\no
\end{eqnarray}
where~\eqref{eq:bound} follows from to the following
Chernoff-Hoeffding bounds, which says that for $n$ i.i.d Bernoulli
random variables $X_j, j=1,\cdots, n$ with mean $\bar{\theta}$,
\begin{eqnarray}\label{eq:chernoffHoeffbound}
\text{Pr}\left\{\left|\frac{\sum X_j}{n}-\bar{\theta}\right|\geq
\epsilon\right\}\leq 2\exp^{-2n\epsilon^2}, \text{ for all
}\epsilon>0.
\end{eqnarray}
To see this, we note that in our case, $X_l(j)$ is the sum of
$Y_l(j)$ i.i.d Bernoulli random variables with parameter
$\theta_l$. On setting
\begin{eqnarray}
n=Y_l(j), \;\text{and}\;\epsilon=\sqrt{\frac{2\ln j}{Y_l(j)}}\no,
\end{eqnarray}
also using the fact that
$$\hat{\theta}_l=\frac{\sum Z_{l}(j)}{Y_l(j)},$$ we
have~\eqref{eq:bound}.

 Similarly, we have
\begin{eqnarray}
&&\hspace{-10mm}\text{Pr}\left\{\Lambda_i(j)\geq
\theta_i+2\sqrt{\frac{2\ln j}{Y_i(j)}}\Bigg|Y_i(j)\geq
m\right\}\no\\&=&\text{Pr}\left\{\hat{\theta}_i\geq
\theta_i+\sqrt{\frac{2\ln j}{Y_i(j)}}\Bigg|Y_i(j)\geq
m\right\}\no\\
&=&\sum\limits_{q=m}^j\text{Pr}\{Y_i(j)=q\}\no\\
&&\hspace{6mm}\text{Pr}\left\{\hat{\theta}_i\geq
\theta_i+\sqrt{\frac{2\ln j}{Y_i(j)}}\Bigg|Y_i(j)\geq
m,Y_i(j)=q\right\}\no\\
&\leq&\sum\limits_{q=1}^j\text{Pr}\left\{\hat{\theta}_i\geq
\theta_i+\sqrt{\frac{2\ln j}{Y_i(j)}}\Bigg|Y_i(j)=q\right\}\no\\
&\leq&2j\exp^{-4\ln j}\no\\ &=& 2j^{-3}.\no
\end{eqnarray}
At the same time, if we set
$$m=\left\lceil\frac{8\ln T}{(\theta_i-\theta_M)^2}\right\rceil,$$
we have for any $1\leq l\leq M$, if $Y_i(j)\geq m$
\begin{eqnarray}
&&\hspace{-18mm}\theta_i+2\sqrt{\frac{2\ln
j}{Y_i(j)}}\leq\theta_i+2\sqrt{\frac{2\ln j}{m}}\no\\
&&\hspace{4mm}\leq\theta_i+(\theta_M-\theta_i)\sqrt{\frac{\ln
j}{\ln T}}<\theta_M\leq \theta_l.\no
\end{eqnarray}
Hence with this $m$, $$\text{Pr}\left\{\theta_l\leq
\theta_i+2\sqrt{\frac{2\ln j}{Y_i(j)}}\Bigg|Y_i(j)\geq
m\right\}=0,$$ for each $1\leq l\leq M$.

Thus,
\begin{eqnarray}
&&\hspace{-10mm}\text{Pr}\left\{\Lambda_{i}(j)\geq
\Lambda_{l}(j)\Big|Y_i(j)\geq m\right\}\no\\&\leq&
\text{Pr}\left\{\Lambda_l(j)\leq \theta_l\Big|Y_i(j)\geq
m\right\}\no\\&&+\text{Pr}\left\{\Lambda_i(j)\geq
\theta_i+2\sqrt{\frac{2\ln j}{Y_i(j)}}\Bigg|Y_i(j)\geq
m\right\}\no\\ &&+\text{Pr}\left\{\theta_l\leq
\theta_i+2\sqrt{\frac{2\ln j}{Y_i(j)}}\Bigg|Y_i(j)\geq
m\right\}\no\\ &\leq& 4j^{-3}.\no
\end{eqnarray}
\begin{eqnarray}\label{eq:inbound}
&&\hspace{-8mm}\mathbb{E}\{Y_i(T)\}\no\\&&\hspace{-8mm}\leq\mathbb{E}\Bigg\{
\sum\limits_{l=1}^M\Bigg\{m+\no\\
&&\hspace{4mm}\sum\limits_{j=\lceil N/M\rceil+m}^T
I\left\{\Lambda_{i}(j)\geq \Lambda_{l}(j)\Big|Y_i(j)\geq
m\right\}\Bigg\}\Bigg\}\no \\
&&\hspace{-8mm}= \sum\limits_{l=1}^M\Bigg\{\left\lceil\frac{8\ln
T}{(\theta_i-\theta_M)^2}\right\rceil+\sum\limits_{j=\lceil
N/M\rceil+m}^T\no\\&&\hspace{4mm}
\mathbb{E}\left\{I\left\{\Lambda_{i}(j)\geq
\Lambda_{l}(j)\Big|Y_i(j)\geq \left\lceil\frac{8\ln
T}{(\theta_i-\theta_M)^2}\right\rceil\right\}\right\}\Bigg\}\no \\
&\leq&M\left\{\left\lceil\frac{8\ln
T}{(\theta_i-\theta_M)^2}\right\rceil+\sum\limits_{j=\lceil
N/M\rceil+m}^T4j^{-3}\right\}\no\\&\sim&O(\ln T),\no
\end{eqnarray}
since
\begin{eqnarray}
\sum\limits_{j=\lceil
N/M\rceil+m}^T4j^{-3}\leq\sum\limits_{j=1}^{\infty}4j^{-3},\no
\end{eqnarray}
and $\sum\limits_{j=1}^{\infty} j^{-3}$ exists.

Hence from~\eqref{eq:inbound}, we have that, for any channel that
is not among the best $M$ channels, the average number of time
slots for which this channel is selected is bounded by $O(\ln T)$.
Thus, the loss is of order $O(\ln T)$.

On the other hand, it has been proved in~\cite{Anantharam:TAC:87}
that for any consistent strategy,
\begin{eqnarray}
\lim\inf\limits_{T\rightarrow\infty} \frac{L(\thetav;\Gamma)}{\ln
T}\geq c_1,\no
\end{eqnarray}
with some constant $c_1$. This completes the proof.
\end{proof}


\section{Conclusions}\label{sec:con}
This work has developed a unified framework for the design and
analysis of cognitive medium access based on the classical bandit
problem. In the single user scenario, our formulation highlights
the tradeoff between exploration and exploitation in cognitive
channel selection. A linear complexity cognitive medium access
algorithm, which is asymptotically optimal as $T\rightarrow
\infty$, has been proposed. The multi-user setting has also been
formulated, as a competitive bandit problem enabling the design of
efficient and game theoretically fair medium access protocols.
Finally, these ideas have been extended to the multi-channel
scenario in which the cognitive user is capable of utilizing
several channels simultaneously.

Our results motivate several interesting directions for future
research. For example, developing optimal medium access strategies
by taking sensing error into consideration and is of practical
significance. Applying other powerful tools from sequential
analysis to design and analyze wireless networks is a promising
research direction.

\end{document}